\documentclass[aps,10pt,tightenlines,eqsecnum,nofootinbib]{revtex4}
\usepackage{amsmath,amscd,amssymb}
\usepackage{color,epsfig}

\def\be{\begin{equation}}
\def\ee{\end{equation}}
\def\bea{\begin{eqnarray}}
\def\eea{\end{eqnarray}}
\def\mult{\mbox{''{\boldmath $\overline{10}$}''}}

\begin{document}

\title{Reply to Cohen's comment on the rotation--vibration
coupling\\ in chiral soliton models}

\author{H. Walliser and H. Weigel}

\affiliation{Fachbereich Physik, Siegen University,
D--57068 Siegen, Germany}

\begin{abstract}
In this short note we summarize the main results of our
paper [hep-ph/0510055] and reply to a recent comment [hep-ph/0511174]
on that paper.

\end{abstract}

\maketitle

\bigskip
\appendix
\setcounter{section}{18}
In a recent comment~\cite{cohen} Cohen criticized our conclusion in 
ref.~\cite{ww} that the rigid rotator approach~(RRA) to generate baryon 
states with non--zero strangeness in chiral soliton models is suitable to 
estimate excitation energies and decay properties of exotic baryons such 
as the $\Theta^+$ pentaquark. Starting point for this criticism is the 
so--called bound state approach (BSA) to chiral soliton models. The BSA 
describes baryons with non--zero strangeness as compound objects of the 
soliton and kaon modes that are treated as harmonic vibrations about the 
soliton. It is well established that the BSA becomes exact in the limit 
that the number of colors, $N_C$ approaches infinity. Cohen's criticism is 
based on the (correct) observation that the excitation energy of the mode 
needed to build the $\Theta^+$ pentaquark does not vanish even in the 
combined limit of large $N_C$ and $m_K\to m_\pi$. Hence rotational and 
vibrational modes do not decouple for pentaquark baryons. Cohen then argues 
that this prevents the introduction of collective coordinates to describe 
these modes as rigid rotations and that the RRA would be inadequate to 
compute physical properties of exotic baryons in large $N_C$ (see also 
refs. [2-7] in ref.~\cite{cohen})\footnote{This (wrong) argument would 
also invalidate the RRA for non--exotic $\mathbf{8}$ and 
$\mathbf{10}$ baryons in the full calculation, where a sizable 
symmetry breaking must be included. Note {\it e.g.\@} that the
excitation energy of the $\Omega(1670)$ is also order $N_C^0$ but even 
larger than that of the $\Theta^+$.}. Conversely, the correct 
conclusion from this observation is that these non--vanishing 
rotation--vibration couplings must be taken into account. This is 
exactly what we did in ref.~\cite{ww}. We found that the correction to 
the RRA estimate of the $\Theta^+$ excitation energy due to the vibrational 
modes is indeed small. For this and other reasons we concluded that the 
$\Theta^+$ may well be considered as a collective excitation of the 
soliton. Here we will back up this conclusion by briefly 
recapitulating the central results of ref.~\cite{ww}.

In the RRA the $SU(3)$ Euler 
angles $\vec{\alpha}$ that parameterize the orientation of the soliton 
in flavor space are introduced as collective coordinates and quantized 
canonically. In the flavor symmetric case the RRA then predicts the 
excitation energy and wave--function of the exotic $\Theta^+$ to be
\be
\omega_\Theta = E_\Theta - E_N = \frac{N_C + 3}{4\, \Theta_K}\, ,
\qquad \langle \vec{\alpha} | \Theta^+  \rangle
\propto D^{\mult}_{(2,0,0),(1,\frac{1}{2},-J_3)}
(\vec{\alpha})\,.
\label{eqR1}
\ee
Form and numerical value of the kaonic moment of inertia, $\Theta_K$, 
depend on the considered model. The baryon wave--functions are Wigner 
$D$-functions of the Euler angles, characterized by the left 
and right quantum numbers $(Y,T,T_3)$ and $(Y_R,J,-J_3)$, respectively,
and the SU(3) representation $\mult$ with $(p,q)=(0,\frac{N_C+3}{2})$ 
for arbitrary $N_C$. Flavor symmetry breaking can straightforwardly 
be included and the exact eigenstates of the 
Hamiltonian ~(4.15)\footnote{The equations in this note 
are labeled (R1), (R2) and (R3), all other numbers refer to formulas 
in ref.~\cite{ww}.} for the collective coordinates
are obtained as linear combinations of states from 
different SU(3) representations. To study the rotation--vibration coupling 
small amplitude fluctuations must be introduced in addition to the 
collective rotations. In ref.~\cite{ww} we have utilized Dirac's
quantization procedure under constraints to quantize these additional 
fluctuations in the subspace that is orthogonal to the rigid rotations 
parameterized by the collective coordinates. This then 
defines the rotation--vibration approach (RVA). An important feature of 
the RVA is that it generates a contribution in the Hamiltonian, 
$H_{\rm int}$ that is linear in these fluctuations. Since $H_{\rm int}$ 
also contains collective coordinate operators it gives rise to a Yukawa 
coupling between the nucleon and its collective excitations. Actually, 
rotation--vibration coupling has been frequently considered in former 
soliton calculations, both in SU(2) and in SU(3) (see \cite{ww} for 
references).

Since the RVA contains collective rotations and orthogonal fluctuations
but the BSA contains fluctuations only, the large $N_C$ correspondence is
such that the fluctuations in the two approaches are equal in the 
subspace orthogonal to the rotations. In the rotational subspace the 
BSA fluctuations must thus correspond to the collective 
rotations of the RRA. In section III and IV of ref.~\cite{ww} we 
therefore have carefully compared the BSA and RRA in the rotational
subspace. Projecting the BSA equation (3.5) onto its rotational subspace 
immediately leads to the criticized eqs.~(3.10) and~(3.11) for the mass 
differences $\omega_\Lambda = E_\Lambda - E_N$ and 
$\omega_\Theta = E_\Theta - E_N$. In Fig.~3 of ref.~\cite{ww} we compared
these mass differences to the excitation energies predicted by the RRA
for arbitrary $N_C$ and $m_K=495{\rm MeV}$. Their equality for large $N_C$
and arbitrary $m_K \not= m_\pi$
unambiguously confirms the above described scenario for the correspondence 
between the BSA fluctuations and the collective excitations.

The central equations of the RVA are the integro--differential eqs.~(5.10) 
for $m_K = m_\pi$ and (7.4) for $m_K \not= m_\pi$. As a matter of fact, 
these equations are fundamental to the RVA and everything else directly 
follows thereof.  We have solved these two equations numerically in order 
to obtain the phase shifts. For completeness we show these phase shifts 
here in an extra figure although they may be easily extracted from 
Figs. 2, 5 and 6 of ref.~\cite{ww}. For $N_C=3$ we notice a sharp and 
pronounced resonance with almost a full $\pi$ jump in the phase shifts.
\begin{figure}[t]
%\vspace{-2mm}
\begin{center}
\begin{tabular}{cc}
\epsfig{figure=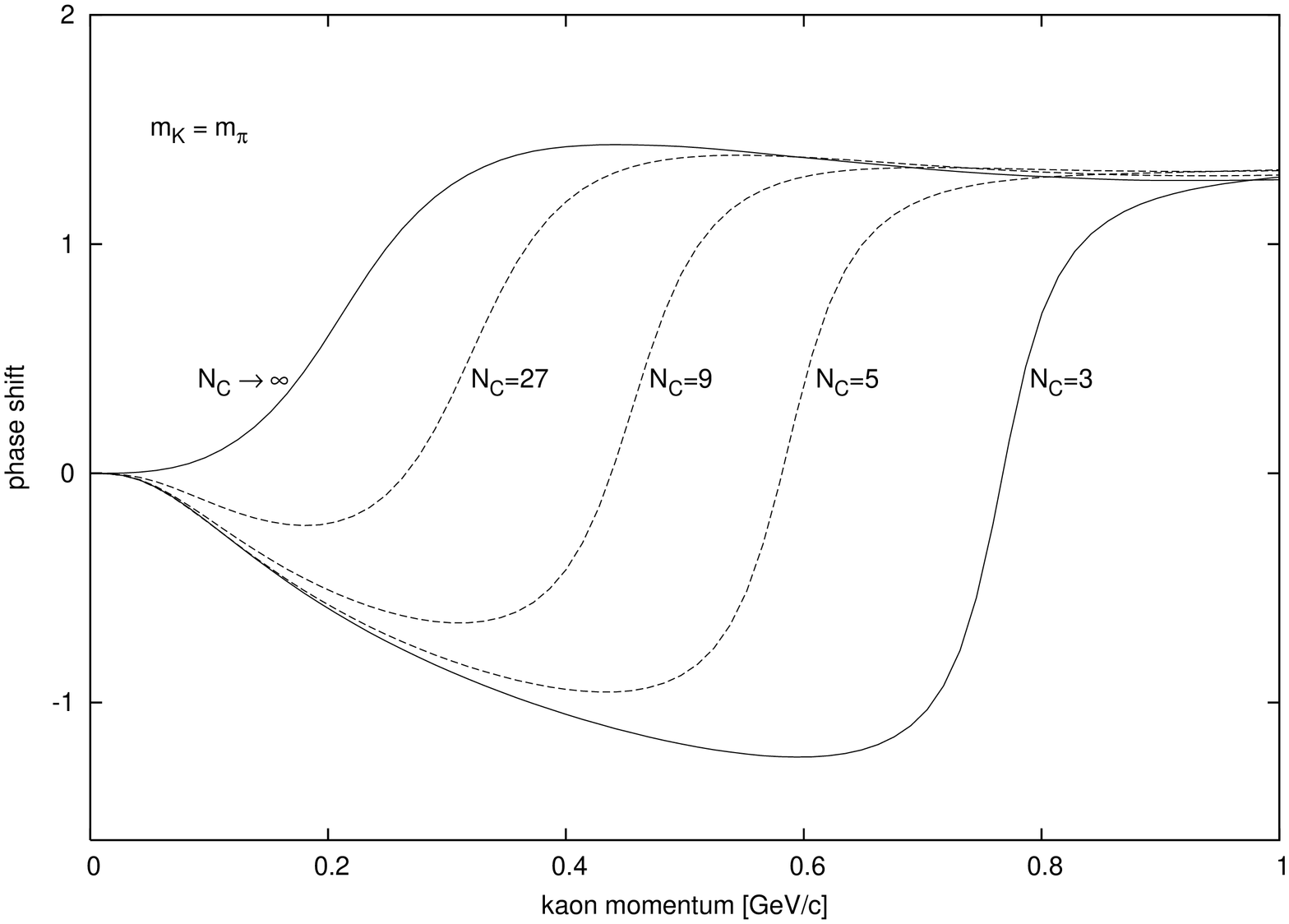,width=4.5cm,angle=270} &
\epsfig{figure=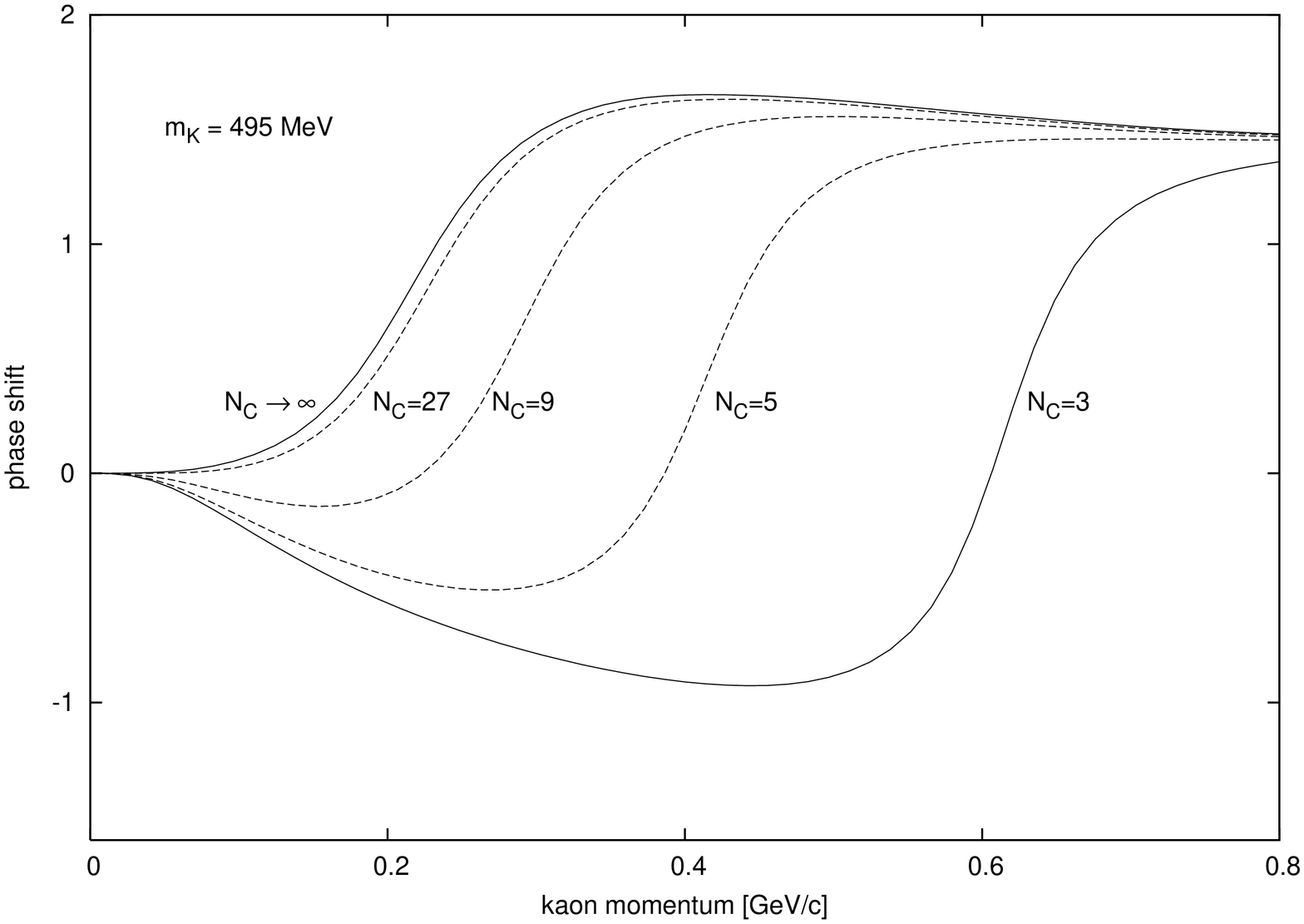,width=4.5cm,angle=270}
\end{tabular}
\protect\caption[]{Full phase shifts as calculated
directly from the RVA equations (5.10) and (7.4)
for various values of $N_C$.
}
\end{center}
%\vspace{-6mm}
\end{figure}
In the RVA the transition matrix element 
$\langle N| H_{\rm int} |\Theta^+\rangle$ between the nucleon 
and the $\Theta^+$ is essential. This matrix element can be 
expressed as a sum of terms that are products of
two factors, (i) a spatial integral over the 
wave--functions of the fluctuations and the soliton profile
and (ii) a collective coordinate matrix element involving the
Wigner $D$--functions of the nucleon and the $\Theta^+$, {\it cf.}
eq.~(\ref{eqR1}). Of course, we have taken configuration
mixing into account in the physical case of $m_K\ne m_\pi$. 

Although our results do not rely on separating background and
resonance phase shifts it is instructive to do so. For
simplicity we consider the SU(3) symmetric case (5.10) and switch
off the $\Lambda$ pole contribution. In the $\Theta^+$
resonance region that contribution is unimportant and in large
$N_C$ it vanishes anyhow if $m_K=m_\pi$~\cite{ww}. Using standard
scattering theory techniques we then find the \emph{exact} and
\emph{unambiguous} relation
\be
\delta(k) = \overline{\delta}(k) + \arctan
\frac{\Gamma_\Theta(\omega_k)/2}{\omega_\Theta - \omega_k 
+ \Delta_\Theta(\omega_k)}\, . 
\label{phase}
\ee
The $N_C$ independent background phase shift $\overline{\delta}(k)$ 
is obtained from (5.10) for vanishing Yukawa coupling. 
Eq.~(\ref{phase}) corroborates that the RRA excitation energy
$\omega_\Theta$ is absolutely essential to reproduce the correct phase
shift within the RVA. The width, $\Gamma_\Theta(\omega_k)$ is proportional 
to the square of the transition matrix element
$\langle N| H_{\rm int} |\Theta^+\rangle$ between the nucleon and the 
$\Theta^+$. The unique resonance contribution arises solely due to the 
Yukawa coupling. It emerges in the standard shape parameterized by 
the width $\Gamma_\Theta$ and the pole shift $\Delta_\Theta$ that are 
listed in eqs.~(6.5) and~(6.6). The collective RRA quantities, 
eq.~(\ref{eqR1}) inevitably enter the computation of 
$\omega_\Theta$, $\Gamma_\Theta$ 
and $\Delta_\Theta$, therewith emphasizing the collective nature of the
$\Theta^+$.  Furthermore these collective coordinate matrix elements induce 
a strong $N_C$ dependence in the resonance contribution. In 
the flavor symmetric case $\langle N| H_{\rm int} |\Theta^+\rangle$ 
contains only a single SU(3) structure. This is in sharp contrast to the 
approaches of refs.~\cite{Diak} that attempt to describe the (potentially) 
small width of the $\Theta^+$ from cancellations between contributions 
from different SU(3) structures. Moreover, the SU(3) structure in
$H_{\rm int}$ is not related to the transition operator for the 
decay $\Delta\to \pi N$. For $N_C=3$ we have calculated a small 
pole shift $\Delta_\Theta=-14{\rm MeV}$. This small number has to be 
contrasted with the RRA excitation energy $\omega_\Theta = 792{\rm MeV}$. 
Obviously, the coupling to the continuum yields a negligible correction to 
the RRA prediction for the excitation energy of the $\Theta^+$. This 
additionally indicates its collective nature.

We have already noted that the BSA is exact for $N_C\to\infty$. Indeed
we have verified that in this limit $\delta(k)$ is identical to the 
BSA phase shift. Nevertheless for $N_C\to\infty$ the separation 
in eq.~(\ref{phase}) still holds and we observe a broad resonance 
hidden by repulsive background phase shifts ({\it cf\@.} Fig.~2 in 
ref.~\cite{ww} for the individual contributions).

Eq.~(\ref{phase}) also applies to the 
$\Delta$ decay in the SU(2) version of the model, where nobody doubts the 
validity of the RRA. Apart from the different transition operator, 
the collective $\Theta^+$ quantities, eq.~(\ref{eqR1}),
are simply replaced by those of the $\Delta$ in the two flavor
model 
\be
\omega_\Delta = E_\Delta - E_N = \frac{3}{2\, \Theta_\pi}\, ,
\qquad \langle \vec{\alpha} | \Delta  \rangle
\propto D^{T=J=\frac{3}{2}}_{T_3,-J_3}(\vec{\alpha})\,,
\ee
where $\Theta_\pi$ is the pionic moment of inertia. A small pole shift 
$\Delta_\Delta$ due to the coupling to the continuum appears also 
there~\cite{Hayashi}. In the large $N_C$ limit width and pole shift 
become sizable for the $\Theta^+$ (cf. Fig. 1) but vanish for the 
$\Delta$ in SU(2). For the $\Delta$ this reflects the above mentioned 
decoupling of rotational and vibrational modes and the fact that the 
$\Delta$ excitation becomes purely collective in that limit. In the 
real world, $N_C=3$, the situation is just reversed, namely width and
pole shift for the $\Theta^+$ are smaller than the corresponding
quantities for the $\Delta$, implying that the collective portion
in the total wave function is even higher for the $\Theta^+$ 
than the $\Delta$. In any case, we may safely conclude
that both excitations, the $\Delta$ \emph{and}
the $\Theta^+$ can reliably be described as collective excitations
of the soliton.

Finally we briefly comment on the $1/N_C$ expansion. Admittedly there is 
an inconsistency which we frankly discussed in chapter V of \cite{ww}. 
Namely, we have selected the leading $N_C$ Yukawa couplings only, but 
treated them to all orders in $N_C$ while we omitted subleading terms. 
This is completely sufficient to investigate the relation between the 
BSA (which does not have subleading terms to begin with) and the RVA.  
Because the leading terms taken into account introduce
already an extreme (for $N_C=3$ diverging) $1/N_C$ dependence
({\it cf.\@} section VI.B of ref.~\cite{ww}) serious doubts concerning the 
applicability of $1/N_C$ expansion methods in the context of exotic 
baryons are in place. There is no reason to expect the subleading 
rotation--vibration couplings to be small. Eventually all possible 
terms would have to be taken into account. This would lead to a 
tremendous computational effort including many Yukawa coupling terms 
into a complicated coupled channel calculation~\cite{Bernd} (the inclusion 
of processes like $K N \longrightarrow K \pi N$ would stand at the very 
end of our wish list). Improvements in that direction probably have to 
wait for a clarification of the experimental situation concerning the 
status of exotic states. 

To summarize, we fully reject the criticism raised in
the comment, ref.~\cite{cohen}, in all points. Moreover,
from the presented argumentation it is obvious that the
exotic $\Theta^+$, alike the non-exotic $\Delta$, is predominantly a
collective soliton excitation. Thus we have to reiterate the 
conclusion drawn in ref.~\cite{ww} that the rigid rotator approach 
is indeed appropriate  in predicting pentaquark masses and properties
in chiral soliton models, in sharp disagreement to the statements
made in ref.~\cite{cohen} (and refs. [2-7] therein)
put forward to discredit this approach to chiral soliton models
in flavor SU(3).

\end{document}